\documentclass[12pt]{article}
\usepackage{amsmath,amssymb,color}
\usepackage{cite,epsf}

\makeatletter \@addtoreset{equation}{section} \makeatother

\usepackage{graphicx,array} 
\newcommand{\be}{\begin{equation}}
\newcommand{\ee}{\end{equation}}
\newcommand{\beq}{\begin{equation}}
\newcommand{\eeq}{\end{equation}}
\newcommand{\bea}{\begin{eqnarray}}
\newcommand{\eea}{\end{eqnarray}}
\newcommand{\nn}{\nonumber}

\newcommand{\ba}{\begin{eqnarray}}
\newcommand{\ea}{\end{eqnarray}}

\begin{document}

\begin{titlepage}
\vspace{10pt}
\hfill
{\large\bf HU-EP-07/26}
\vspace{20mm} 
\begin{center}

{\Large\bf  On the internal space dependence of\\[2mm] 
the static quark-antiquark potential\\[3mm] in ${\cal N}=4$ SYM  
plasma wind  }

\vspace{40pt}

{\large  Harald Dorn, Thanh Hai Ngo }

\vspace{20pt}

{\it Humboldt--Universit\"at zu Berlin, Institut f\"ur Physik\\ Newtonstr. 15, D--12489 Berlin}\\ [4mm]

\vspace{10pt}

\centerline{\tt dorn@physik.hu-berlin.de, ngo@physik.hu-berlin.de}

\vspace{40pt}
\end{center}

\centerline{{\bf{Abstract}}}
\vspace{15pt}
\noindent
We study the effect of the relative $S^5$-angle of a quark and an antiquark 
on their static potential and the related screening length in a strongly
coupled moving ${\cal N}=4$ SYM plasma. The large velocity scaling
law for the screening length holds for any relative $S^5$-angle
$\theta$. However, the velocity independent 
prefactor $Z$ strongly depends on $\theta$. For comparison with QCD we propose
to average $Z$ over all relative orientations on $S^5$. This generates
a suppression factor relative to the case $\theta =0$.

\end{titlepage}
\newpage
\noindent
{\large\bf Introduction}\\[2mm]
\noindent
Part of the AdS/CFT dictionary concerns the mapping of Wilson loops in the 
gauge 
theory on the boundary to the string partition function in the bulk with the
condition that the string world sheet approaches the Wilson loop contour
at the boundary \cite{Mal98}. In the prototype case of ${\cal N}=4 $ super
Yang-Mills theory dual to type IIB string theory in $AdS_5\times S^5$, 
specifying the Wilson loop contour implies both fixing the contour in
Minkowski space as well as a contour in $S^5$. This contour in internal space 
parametrises the coupling to the scalar partners of the gauge field bosons.
In the construction of BPS Wilson loops in ${\cal N}=4 $ SYM,
which globally preserve part of the supersymmetry, a variation of the
$S^5$-orientation along the contour turns out to be crucial \cite{Mal98,
  Zarembo, Drukker}.

Along another line of developments, the  AdS/CFT correspondence for Wilson
loops has been widely used to calculate the static quark-antiquark potential 
via  string world
sheets in modified 10-dimensional backgrounds which break all or part of
supersymmetries and are expected to model a gauge theory setup closer to
QCD. 
 Since there is no use for the internal space degree of freedom (related to $S^5$ 
or its modifications) at the end in QCD anyway,  
most of these calculations keep the internal space orientation of the Wilson loop fixed
along the rectangular contour relevant for the evaluation of the potential
. There seem to exist only few papers discussing the
effect of the {\it relative} internal space orientation of two static sources
(quark-antiquark) on the resulting potential 
\cite{Mal98, Dorn1, Dorn2, Sfetsos05}. 

Recently, a lot of activities have been devoted to understand the heavy ion RHIC 
data which probe QCD at strong coupling and high temperature. This is just
a parameter regime inviting for an application of AdS/CFT techniques. In this
letter we will comment on the calculation of the quark-antiquark potential,
the related screening length and the jet quenching parameter via AdS/CFT
from strings in a boosted AdS Schwarzschild black hole background
 \cite{Liu, guijosa, Sfetsos06}.   

It has been argued in ref.\cite{Liu} that these quantities in  ${\cal N}=4 $
SYM and QCD share some universality properties and hence the string
calculations in the boosted AdS Schwarzschild black hole background will tell
us something about QCD under these extreme conditions. In particular 
the screening length scales at large velocities for the boost 
as $\ell_{max}=Z\cdot(1-v^2)^{1/4}$. 
 Here the question naturally arises as to whether this conjectured universality holds within ${\cal N}=4 $
SYM itself 
if a degree of freedom is switched on, which so far has not been
taken into account in this context. 

 For such an additional degree of freedom,
we have in mind the  angle $\theta$
between the  $S^5$-positions of the quark and the antiquark. We can confirm the
universality of the $ (1-v^2)^{1/4}$ behaviour and will find a
$\theta$-dependence in the prefactor $Z$ monotonically decreasing from the already
known value for coinciding $S^5$-positions to zero for antipodal positions
$\theta =\pi$.

The discussion of universality in \cite{Liu}  also contains 
a discussion of model dependence. For the jet quenching parameter  it is argued that
the ratio of its value for QCD to that for  ${\cal N}=4 $ SYM is given by the
square root of the ratio of the number of degrees of freedom. 
It seems 
natural  to apply the same reasoning to the $v$-independent
prefactor $Z$ in the large $v$ scaling law. Since QCD has less degrees of
freedom, this leads to a decrease of the prefactor.

We propose to consider also 
another procedure to care for
the reduction of degrees of freedom in going from ${\cal N}=4 $ SYM to QCD.
Since in QCD there is no use for the internal space degrees of freedom,
we average the $\theta$-dependent prefactor $Z$ over all possible 
$S^5$-positions of the quark relative to the antiquark. This will give us
a suppression factor 0.78 in reference to  ${\cal N}=4 $ SYM with coinciding
positions of quark and antiquark. 

In a last paragraph we discuss the $\theta$ and $\ell$-dependence of the
$q$-$\bar q$ potential in ${\cal N}=4 $ SYM and comment on its concavity
property and the stability of the related string configurations.\\[5mm]
{\large\bf General setup and screening length}\\[2mm] 
\noindent
The 10-dimensional metric for a black hole with Hawking temperature $T_H$ 
boosted with velocity $v$ in $x_3$-direction in $AdS_5\times S^5$ with radius 
$R$ as used in \cite{Liu} is
\be
        ds^2 = - A dt^2 - 2B dt dx_3 + C dx_3^2 + \frac{r^2}{R^2} (dx_1^2 + dx_2
^2)+
        \frac{1}{f}dr^2 + R^2~d\Omega _5^2~,
        \label{BoostedAdSBlack}
\ee
with
\bea
f&=& \frac{r^2 }{ R^2} \left(1 - \frac{r_0
^4 }{ r^4} \right)~,~~~~~~~~~~T_H = \frac{r_0}{\pi R^2}~, \\
A& =& \frac{r^2 }{R^2}  - \frac{r_0^4\gamma ^2}{r^2 R^2} \ , \quad B =
\frac{r_0^4\gamma ^2 v}{r^2 R^2}\ ,\quad
        C = \frac{r^2}{R^2}  + \frac{r_0^4\gamma ^2v^2}{r^2R^2} \ ,\nn
              \label{ABC}
\eea
and $\gamma$ introduced as a short hand for
\be
\gamma = (1-v^2)^{-1/2}~.
\label{gamma}
\ee
The static quark-antiquark potential $V(L,\theta ,v)$, depending
on the distance $L$ of $q$ and $\bar q$ in space, the angle $\theta$ 
between the positions of $q$ and $\bar q$ in the internal $S^5$ and the
velocity $v$ can be read off in the limit $T\rightarrow\infty$ from 
the Wilson loop for a rectangle having extension $T$ in time direction and 
extension $L$ in $x_1$ direction. By  choosing the  $q$-$\bar q$ separation in
$x_1$ direction we 
concentrate on the case of a boost orthogonal to the  $q$-$\bar q$ dipole.
The timelike edges of the rectangle represent the worldlines of the static
$q$ and $\bar q$ respectively, we assume on both of these edges constant
position in $S^5$, but allow a nonzero angle $\theta$ between both 
static positions.

Then the Nambu-Goto action with the induced metric for a string world sheet,
approaching on the boundary of $AdS$ ($r\rightarrow\infty$) the just
discussed rectangle, turns out to be (as usual using translation invariance
in time for large $T$)
\be
S~=~\frac{r_0T}{2\pi\alpha '}~\int
_{-\ell/2}^{\ell/2}dx~\sqrt{(y^4-\gamma^2)\left (1+ \frac{y'^2}{y^4-1}+\frac
{\vartheta '^2}{y^2}\right )}.\label{S}
\ee 
Here we used the dimensionless coordinates $y$, $x$ and the rescaled 
$q$-$\bar q$-separation $\ell$
\be
y=\frac{r}{r_0}~,~~~x=\frac{x_1r_0}{R^2}~,~~~~~\ell =\pi T_HL~.
\label{rescaling}
\ee
The prime denotes differentiation with respect to $x$, and $\vartheta$ is
the $x$-dependent angle on the great circle connecting the $S^5$ position
of $q$ and $\bar q$. The boundary conditions on $y(x)$ and $\vartheta (x)$
are
\be
y(\pm \ell/2)=\infty ~,~~~~~~\vartheta (\pm \ell/2)=\pm ~ \theta/2~.
\label{bc}
\ee
Since the integrand in (\ref{S}) does not depend explicitely on $x$
and is independent of $\vartheta$, there are two conserved quantities
$\epsilon$ and $j$
\bea
        \epsilon &=&
        \sqrt{\frac{y^4-\gamma^2}{1+\frac{y'^2}{y^4-1}+\frac{{\vartheta'} ^2}
{y^2}}}  ~~~=~~~
        \sqrt{\frac{y_c^2\left(y_c^4-\gamma^2\right)}{y_c^2+
{\vartheta'}_c ^2}}~,    \nn \\
j&=&\epsilon \cdot \frac{\vartheta'}{y^2} 
       ~~~~~~ =~~~              {\vartheta'}_c ~\sqrt{\frac{\left(y_c^4-\gamma^2\right)}{y_c^2\left(y_c^2+
{\vartheta'}_c ^2\right)}}\ .
\label{conservation}
\eea
The rightmost sides express these conserved quantities in terms of geometric
characteristics of the string world sheet: $y_c$ the minimal $y$ value realized
for symmetry reasons at $x=0$ where $y'=0$, and $\vartheta'_c=\vartheta'(0)$.  

The integration of (\ref{conservation}) yields ($w=y_c/y$)
\bea
\theta &=&2~\sqrt{1-h}~\int
_0^1\frac{dw}{\sqrt{(1-kw^4/\gamma ^2)(1-w^2)(1+hw^2)}}~,\nn\\
\ell &=& \frac{2k^{1/4}}{\gamma ^{1/2}}~\sqrt{h-k}~\int
_0^1\frac{w^2~dw}{\sqrt{(1-kw^4/\gamma ^2)(1-w^2)(1+hw^2)}}~,\label{theta-l}
\eea
where we switched from $y_c,~\vartheta '_c$ parametrising the 
conserved quantities in (\ref{conservation}) to $h$ and $k$ defined by
\be
h~=~\frac{y_c^6+\gamma ^2{\vartheta'}_c^2}{y_c^6+y_c^4{\vartheta'}_c^2}~,~~~~~~
k~=~\frac{\gamma ^2}{y_c^4}~.\label{hk}
\ee
Real $\epsilon$ and $j$ in (\ref{conservation}) require $y_c^4\geq \gamma^2$,
i.e. $k\in(0,1]$, which together with reality for $\theta$ and $\ell$
in (\ref{theta-l}) constrains the parameters $h$ and $k$ to
\be
0<k\leq h\leq 1 ~.\label{hk-interval}
\ee
Expanding the first factor under the square root in the integrals for large
$\gamma $ one gets a representation in terms of elliptic integrals
\be
\theta =2~\sqrt{1-h}~\Big \{K(-h)+\frac{k}{6h^2\gamma
  ^2}\Big (2(h-1)E(-h)+(2-h)K(-h)\Big )+{\cal O}(\gamma ^{-4})\Big \}\label{exptheta}~,
\ee
\bea
\ell~=~\frac{2k^{1/4}\sqrt{h-k}}{\gamma ^{1/2}}~\Big \{\frac{E(-h)-K(-h)}{h}
&+&\frac{k}{30h^3\gamma ^2}~\Big ((8+h(8h-7))E(-h)\label{expl}\\
&+&(h(3-4h)-8)K(-h)\Big )+{\cal O}(\gamma
^{-4})\Big \}~.\nn 
\eea
In leading large $\gamma$ approximation the relative  $S^5$-angle $\theta$
depends on $h$ only and, in addition, the $\theta\leftrightarrow h$ relation
is one to one with $\theta(0)=\pi$ and $\theta(1) = 0$. This
simplifies the further analysis considerably, and we will restrict ourselves
to this level of approximation throughout the rest of the paper. 

For fixed $\theta$, i.e. fixed $h$, the rescaled dimensionless $q$-$\bar q$
separation $\ell$ depends on the remaining conserved quantity $k$ 
via the trivial $k^{1/4}\sqrt{h-k}$ factor taking into account
the constraint (\ref{hk-interval}). 
This defines a maximal value $\ell _{max}=\ell(k=h/3)$ which
plays the role of a screening length, since for larger $q$-$\bar q$
separation one obviously finds no solution of the stationarity condition
for the Nambu-Goto action (\ref{S})
\bea
\ell_{max}&=&Z(\theta)\cdot\gamma ^{-1/2}~+~{\cal O}(\gamma ^{-5/2})~,\nn\\
Z(\theta)&=&\frac{2\sqrt{2}}{3^{3/4}}~\frac{E(-h)-K(-h)}{(h(\theta ))^{1/4}}~.
\label{lmax}
\eea
The large velocity scaling $\ell_{max}\propto \gamma ^{-1/2}$ holds
for all $\theta \in [0,\pi )$. The prefactor $Z(\theta)$ monotonously
decreases from $Z(0)=0.7433$, known already from \cite{Liu}, to $Z(\pi)=0$,
see also fig.1. For $\ell <\ell_{max}$ there are two solutions to the
stationarity condition, one for $k<h/3$, the ``short string'' and
one for $k>h/3$, the ``long string''.
\begin{figure}[t]
\label{fig1}
\begin{center}
\vskip -1 cm 
\begin{tabular}{cc}
\includegraphics[height=4.8cm]{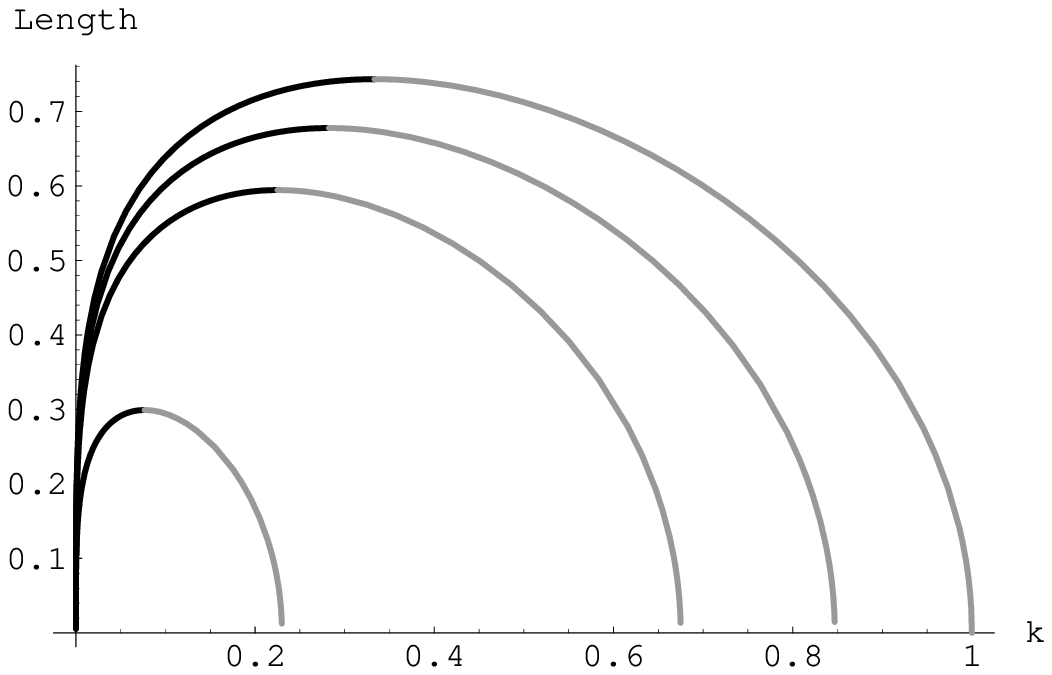} & 
\includegraphics[height=4.8cm]{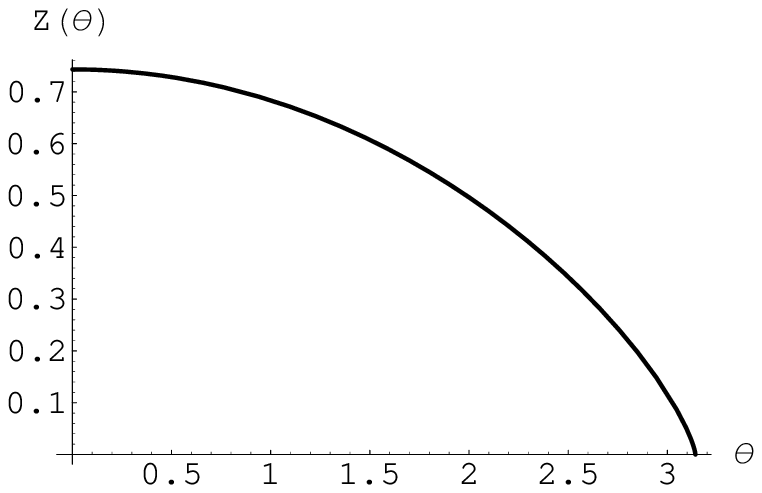}
        \\(a) & (b) 
\end{tabular}
\end{center}
\vskip -.5 cm 
\caption{\textit{(a) The quark antiquark separation $\ell$, in units of $(\pi T_H)^{-1}$, as a function of $k=\frac{\gamma^2}{y_c^4}$ with $\theta = \left\{0,\frac{\pi}{3},\frac{\pi}{2},\frac{5\pi}{6}\right\}$ from the top to bottom respectively. The dark lines refer to the "short strings" while the gray lines to the "long strings".\newline 
(b) The prefactor Z as a function of the angle $\theta$.
}}
\end{figure}

To get out of this analysis something which, assuming some kind
of universality, should be compared to QCD, we propose to average
$Z(\theta)$ over all $\theta$ with a weight $\omega(\theta)$ 
given by the volume per $\theta $ on $\Omega _5$ divided by the total $\Omega
_5$ volume 
\be
\omega (\theta )=\frac{vol\Omega_4}{vol\Omega_5}~\sin ^4\theta
~=~\frac{8}{3\pi}~\sin^4\theta~.
\label{weight}
\ee
Then the average
\be
\overline{Z(\theta)}~=~\int _0^{\pi}Z(\theta)~\omega(\theta )~d\theta
\label{average}
\ee 
using (\ref{exptheta}) can be expressed as an integral over $h$ and evaluated
numerically. The result is $\overline{Z}~=~0.5797$ and gives a suppression
factor $\overline{Z}/Z(0)~=~0.7797$ relative to ref. \cite{Liu}. We expect a
similar averaging to make sense also for the jet quenching parameter.
\\[2mm]
{\large\bf Comments on the potential}\\[2mm] 
\noindent
Via AdS/CFT and the standard relation to the Wilson loop the static $q$-$\bar q$
potential for large 't Hooft coupling $\lambda$ is given by
\be
V(L,\theta,v)~=~S_{stat}/T~,
\ee
where $S_{stat}$ is the Nambu-Goto action (\ref{S}) taken on shell, and one has
to identify $\sqrt{\lambda}=R^2/\alpha '$. Then using (\ref{conservation})
one gets
\be
V~=~\frac{T_H\sqrt{\lambda}}{\epsilon}~\int _0^{\ell/2}(y^4-\gamma
^2)~dx~=T_H\sqrt{\lambda}~\int _{y_c}^{\infty}\frac{(y^4-\gamma
  ^2)~dy}{\sqrt{(y^4-1) (y^4-\gamma ^2-\epsilon ^2-j^2y^2)}}~.
\label{V1}
\ee  
The integral is divergent at $y\rightarrow\infty$. We adopt the usual
procedure to subtract two times the action for a string worldsheet 
 located at fixed $x$ and stretching along $y$ from the horizon to infinity. 
Expressing in addition $\epsilon$ and $\gamma$ via (\ref{conservation}),(\ref{hk}) by
$h$ and $k$ and transforming the integration variable by $y=y_c/w$ we get
\be
\frac{V_{ren}}{\lambda ^{1/2}T_H}=\left\{1+\frac{\gamma^{\frac{1}{2}}}{k^{\frac{1}{4}}}\left (-1+\int
_0^1\Big (\frac{1-kw^4}{\sqrt{(1-\frac{kw^4}{\gamma ^2})(1-w^2)(1+hw^2)}}-1\Big
  )\frac{dw}{w^2}\Big )\right )\right\}.
\ee  
As above in the integrals for $\ell$ and $\theta $, the expansion for large
$\gamma$ allows a representation in terms of elliptic integrals
\be
V_{ren}(L,\theta,v)~=~T_H\lambda^{1/2}\left (1
+\frac{\gamma ^{1/2}(h+k)}{hk^{1/4}}\Big ( K(-h)-E(-h)\Big )+{\cal O}(\gamma
^{-3/2} )\right )~.\label{pot}
\ee

A graphical representation of  this leading contribution to  the potential as a function of the
$q$-$\bar q$ separation $\ell$  for various values of the velocity $v$ and
the $S^5$ angle $\theta$ can be generated by fixing $h(\theta)$
and using (\ref{expl}) and (\ref{pot}) for parametric plots with $k$ as
parameter, see fig.2 and fig.3a.

From the field theoretical point of view the potential 
has to be concave and monotonically growing with  $L$. This follows from
Osterwalder-Schrader positivity of the Euclidean version of the theory
\cite{Bachas}. A generalisation of this rigorous result to the present case,
i.e. including  the relative $S^5$-orientation,  
has been given in \cite{Dorn2}: concavity and monotonic growth in $\ell$
at each fixed $\theta$,  $V(L,\theta)\geq V(L,0)$ and, on a more conjectural
level, concavity in all directions of the $(L,\theta)$-plane. 

Using
(\ref{expl}), (\ref{pot}) one gets
\be
\left (\frac{\partial V_{ren}}{\partial \ell}\right
)_{\theta}~=~\frac{\gamma\sqrt{h-k}}{2\sqrt{k}}~,~~~
\left (\frac{\partial ^2V_{ren}}{\partial \ell ^2}\right
)_{\theta}~=~\frac{h^2\gamma ^{3/2}}{2(h-3k)k^{3/4}(K(-h)-E(-h))}~.
\label{concavity}
\ee 
With (\ref{hk-interval}) and $K(-h)-E(-h)<0~,~\forall h\in (0,1]$ we see that,
while monotony is always realized, concavity holds on the short string
branch ($k<h/3$) only. This perfectly fits with the stability analysis 
on the string side, where it has been shown that the short (long) string
branch is stable (unstable) with respect to small fluctuations
\cite{Gubser,Sfetsos-stab}.  We did not analytically check the extended
concavity in the $(L,\theta)$-plane, but the graph in fig.2a suggests
its validity.

As seen from fig.2, for large enough $\theta$, part of the short string branch 
reaches positive values for the potential. Due to our renormalisation
prescription this implies a metastable situation  \cite{guijosa,
  Sfetsos-stab}.   In spite of the stability with respect
to small fluctuations, the configuration of two separated worldsheets
located at fixed $x$ and stretching along $y$ from the horizon to infinity
would be favoured. 

At first sight this metastability in the neighbourhood of 
 $\ell_{max}$  could obstruct our averaging proposal
for the screening length advocated above. However, another look 
at fig.3a indicates a weakening of this effect for increasing
$\gamma $,
such that the critical 
value for $\theta$ is driven to $\pi$
for $\gamma\rightarrow\infty$, see fig.3b. 
Since after all our discussion of the screening length
concerns its leading behaviour for large $\gamma$, no objection
to an averaging over the whole $S^5$ remains.
\begin{figure}[t]
\label{fig2}
\begin{center}
\vskip -1 cm 
\begin{tabular}{cc}
\includegraphics [height=5.2cm]{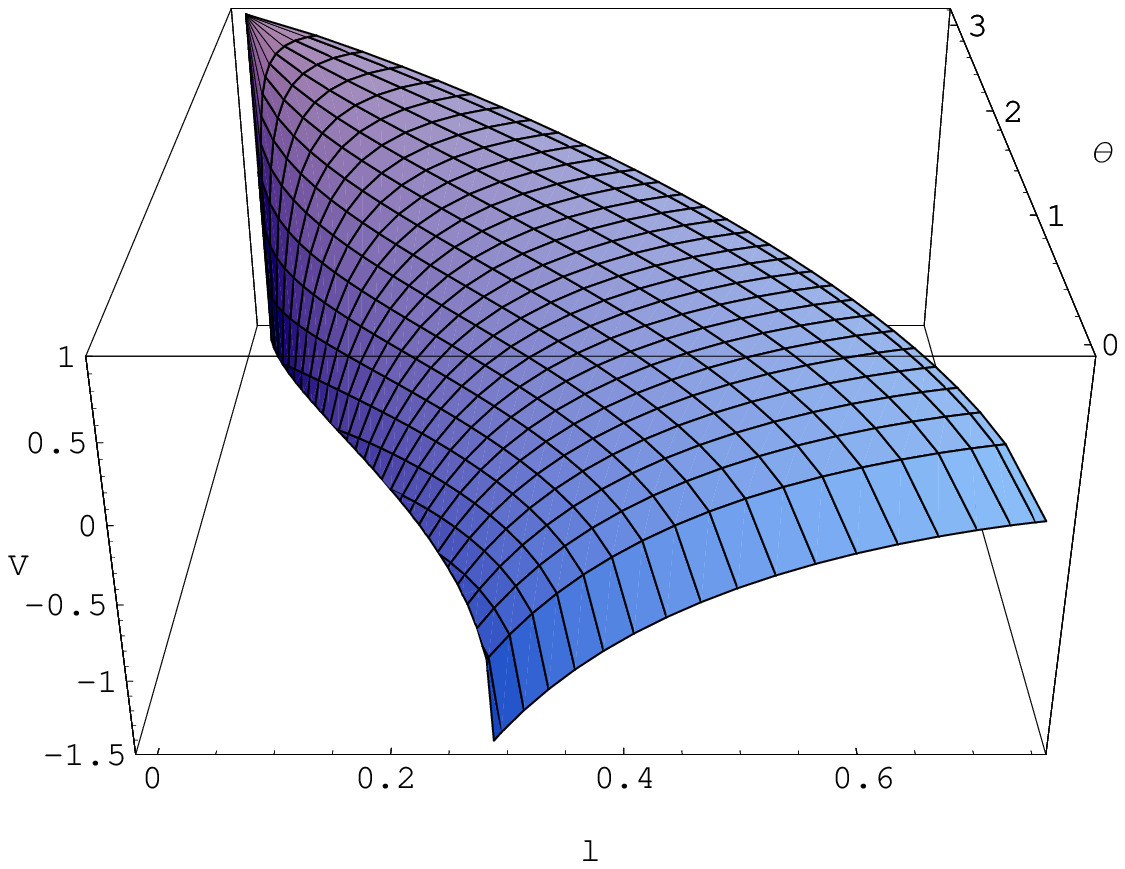} &
\includegraphics[height=4.8cm]{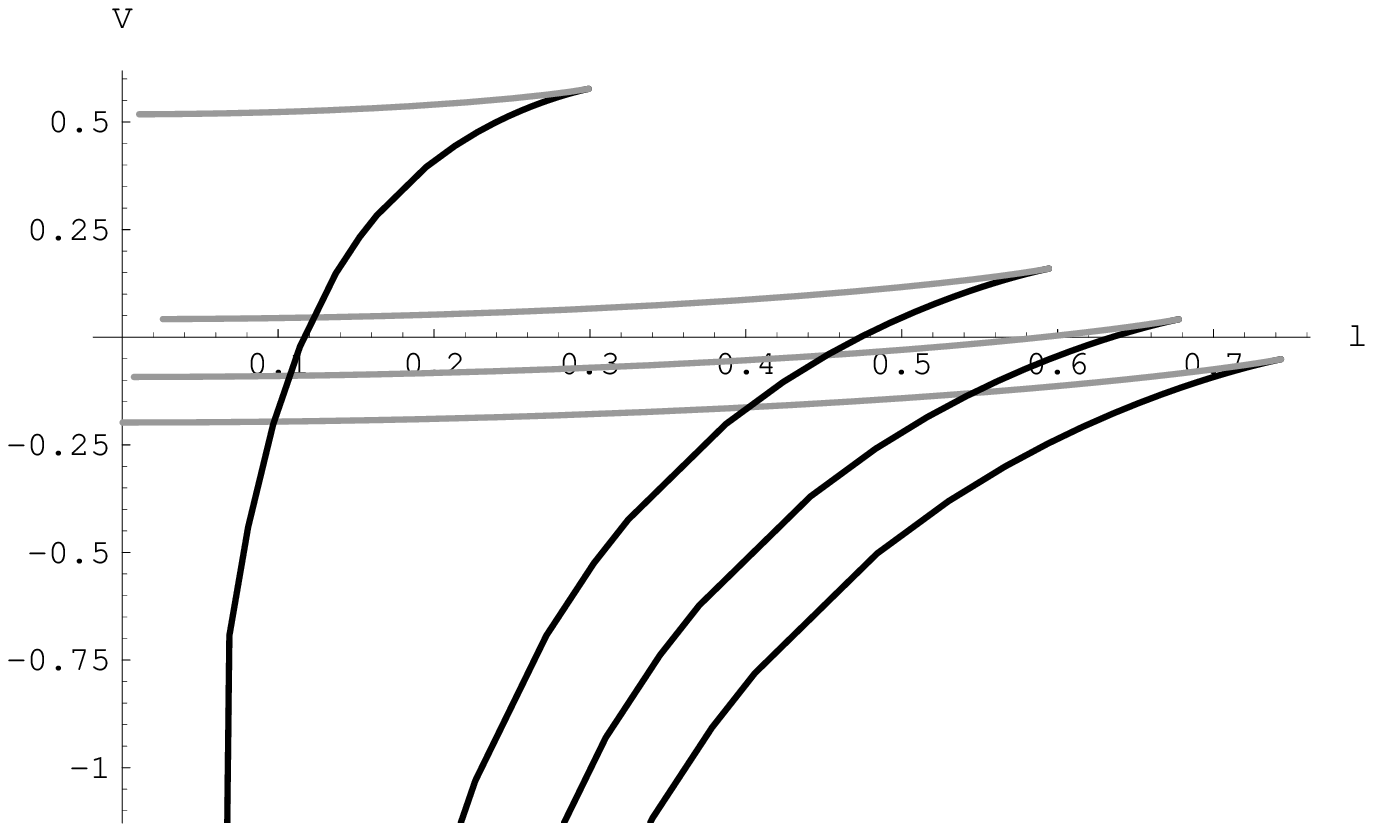}
\\
(a) & (b) 
\end{tabular}
\end{center}
\vskip -.5 cm 
\caption{\textit{ (a)  Leading order of the  potential  $V_{ren}$ 
    for the short string branch, in units\newline $~$\hfill
  of  $ \lambda ^{\frac{1}{2}} T_H $, as a
  function of  separation  $\ell $, in units of $\frac{1}{\pi T_H}$, and 
  angle   $\theta $ for $\gamma=1$.\newline
(b)  Leading $V_{ren}$  for $\gamma=1$ and $\theta = \left\{
  \frac{5\pi}{6},\frac{\pi}{2},\frac{\pi}{3},0\right\}$ from  top to bottom
   respectively. \newline $~~~~$
The dark lines refer to   "short strings"   while the
  gray lines to   "long strings".}}  
\end{figure}
\begin{figure}[t]
\label{fig3}
\begin{center}
\begin{tabular}{cc}
\includegraphics [height=4.8cm]{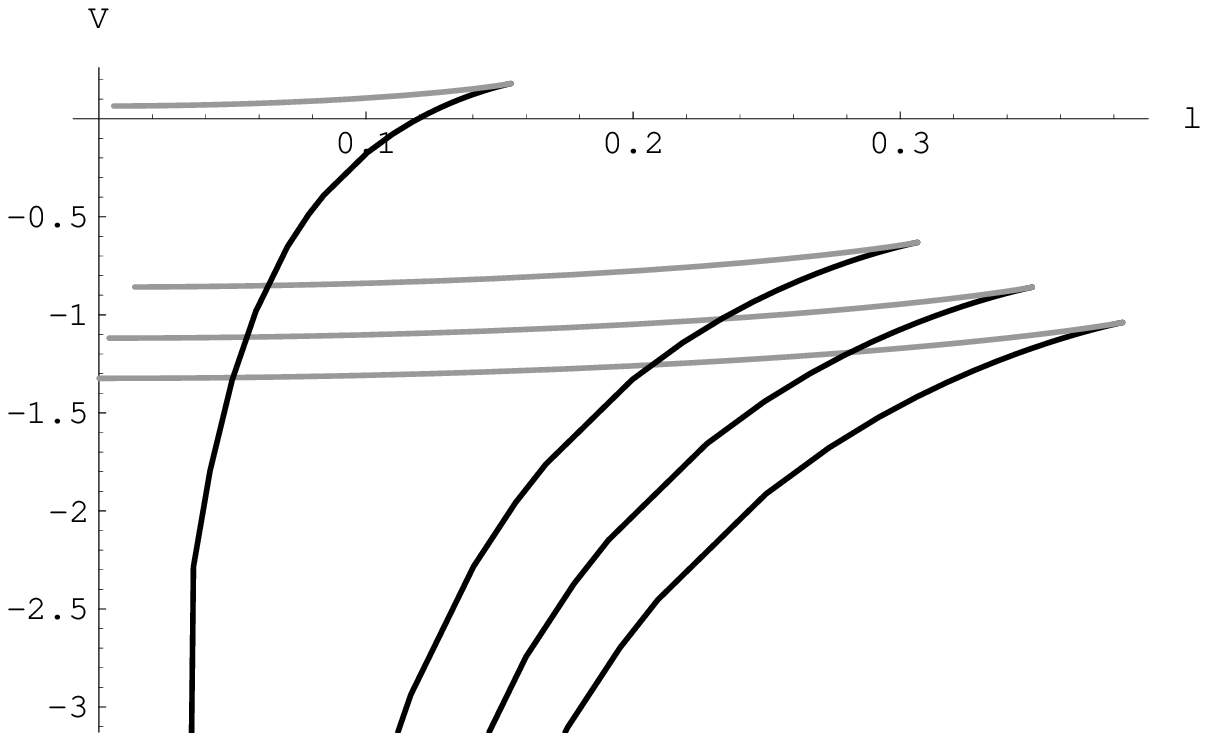} &
\includegraphics [height=4.8cm]{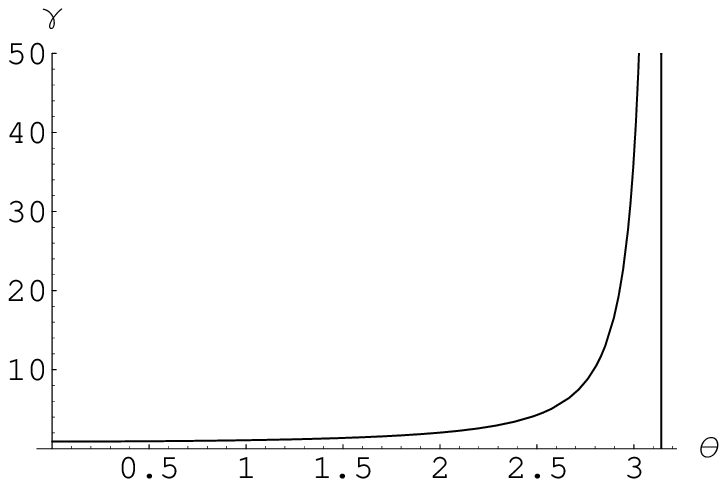}
\\
(a) & (b) 
\end{tabular}
\end{center}
\vskip -.5 cm 
\caption{\textit{(a) Basically the same representation as fig.2b but for $\gamma = cosh(2)$.\newline
(b) Stability analysis represented on the $\gamma$-$\theta$-plane, the short
string configuration \newline $~~~~$ is stable (metastable)  up to  $\ell_{max}$  in the regions above (below) the curve.}}
\end{figure}

 Note that we just discussed the leading large velocity contribution to
  $V_{ren}$. For the full potential metastability appears even at
$\theta =0$ for values of $v<0.45$ i.e. $\gamma <1.12 $ \cite{guijosa}. 
\\[5mm]
\noindent
{\bf Acknowledgement}\\[2mm] 
\noindent
This work has been supported in part by the German Science Foundation (DFG)
grant DO 447/4-1. H.D. thanks K. Sfetsos for a useful discussion.


\begin{thebibliography}{99}

\bibitem{Mal98}
  J.~M.~Maldacena,
  Phys.\ Rev.\ Lett.\  {\bf 80} (1998) 4859
  [arXiv:hep-th/9803002],\\
S.~J.~Rey and J.~T.~Yee,
  Eur.\ Phys.\ J.\  C {\bf 22} (2001) 379
  [arXiv:hep-th/9803001].

\bibitem{Zarembo}
  K.~Zarembo,
  Nucl.\ Phys.\  B {\bf 643} (2002) 157
  [arXiv:hep-th/0205160].

\bibitem{Drukker}
  N.~Drukker,
  JHEP {\bf 0609} (2006) 004
  [arXiv:hep-th/0605151],\\
  N.~Drukker, S.~Giombi, R.~Ricci and D.~Trancanelli,
  ``More supersymmetric Wilson loops,''
  arXiv:0704.2237 [hep-th].

\bibitem{Dorn1}
 H.~Dorn and H.~J.~Otto,
  JHEP {\bf 9809} (1998) 021
  [arXiv:hep-th/9807093].

\bibitem{Dorn2}
 H.~Dorn and V.~D.~Pershin,
  Phys.\ Lett.\  B {\bf 461} (1999) 338
  [arXiv:hep-th/9906073].

\bibitem{Sfetsos05}
 R.~Hernandez, K.~Sfetsos and D.~Zoakos,
  JHEP {\bf 0603} (2006) 069
  [arXiv:hep-th/0510132].

\bibitem{Liu}
 H.~Liu, K.~Rajagopal and U.~A.~Wiedemann,
  Phys.\ Rev.\ Lett.\  {\bf 98} (2007) 182301
  [arXiv:hep-ph/0607062].\\
  H.~Liu, K.~Rajagopal and U.~A.~Wiedemann,
  JHEP {\bf 0703} (2007) 066
  [arXiv:hep-ph/0612168].

\bibitem{guijosa}
 
M.~Chernicoff, J.~A.~Garcia and A.~Guijosa,
  JHEP {\bf 0609} (2006) 068
  [arXiv:hep-th/0607089].
 

\bibitem{Sfetsos06}
 S.~D.~Avramis and K.~Sfetsos,
  JHEP {\bf 0701} (2007) 065
  [arXiv:hep-th/0606190],\\
S.~D.~Avramis, K.~Sfetsos and D.~Zoakos,
  Phys.\ Rev.\  D {\bf 75} (2007) 025009
  [arXiv:hep-th/0609079],\\
P.~C.~Argyres, M.~Edalati and J.~F.~Vazquez-Poritz,
  JHEP {\bf 0704} (2007) 049
  [arXiv:hep-th/0612157].

\bibitem{Bachas}
  C.~Bachas,
  Phys.\ Rev.\  D {\bf 33} (1986) 2723.

\bibitem{Gubser}
  J.~J.~Friess, S.~S.~Gubser, G.~Michalogiorgakis and S.~S.~Pufu,
  JHEP {\bf 0704} (2007) 079
  [arXiv:hep-th/0609137].

\bibitem{Sfetsos-stab}
  S.~D.~Avramis, K.~Sfetsos and K.~Siampos,
  Nucl.\ Phys.\  B {\bf 769} (2007) 44
  [arXiv:hep-th/0612139].


\end{thebibliography}
\end{document}